\documentclass[]{aastex63}%[linenumbers]{aastex63}
\usepackage{graphicx}
\usepackage{subfigure}
\usepackage{color}
\usepackage{amsmath}
\usepackage{pifont}
\usepackage{multirow}
\usepackage{threeparttable}
\usepackage{colortbl}
\usepackage{epstopdf}
\shorttitle{Event Rate of FRB from BNS-Mergers}
\shortauthors{Chen et al.}
\newcommand{\MyFigA}{\ref{MyFigA}}
\newcommand{\MyFigB}{\ref{MyFigB}}

\begin{document}
\title{Event Rate of Fast Radio Burst from Binary Neutron-star Mergers}
\correspondingauthor{Da-Bin Lin}
\email{lindabin@gxu.edu.cn}
\author[0000-0002-0926-5406]{Zhi-Lin Chen}
\affil{Guangxi Key Laboratory for Relativistic Astrophysics, School of Physical Science and Technology, Guangxi University, \\Nanning 530004, China}
\author[0000-0002-6442-7850]{Rui-Chong Hu}
\affil{Guangxi Key Laboratory for Relativistic Astrophysics, School of Physical Science and Technology, Guangxi University, \\Nanning 530004, China}
\author[0000-0003-1474-293X]{Da-Bin Lin}
\affil{Guangxi Key Laboratory for Relativistic Astrophysics, School of Physical Science and Technology, Guangxi University, \\Nanning 530004, China}
\author[0000-0002-7044-733X]{En-Wei Liang}
\affil{Guangxi Key Laboratory for Relativistic Astrophysics, School of Physical Science and Technology, Guangxi University, \\Nanning 530004, China}
%\date{Received 30 Dec. 2022 / Accepted 02 July 2023}
\begin{abstract}
It is proposed that one-off fast radio burst (FRB) with periodic structures
may be produced during the inspiral phase of a binary neutron-star (BNS) merger.
In this paper, we study the event rate of such kind of FRB.
We first investigate the properties of two one-off FRBs with periodic structures
(i.e., FRB~20191221A and FRB~20210213A) in this scenario,
by assuming the fast magnetosonic wave is responsible for their radio emission.
For the luminosities and periods of these bursts,
it is found that the pre-merger BNS with magnetic field strength $B\gtrsim 10^{12}\,{\rm Gs}$ is required.
This is relatively high compared with that of the most of the BNSs observed in our Galaxy,
of which the magnetic field is around $10^{9}\,{\rm Gs}$.
Since the observed BNSs in our Galaxy are the binaries without suffering merger,
a credited event rate of BNS-merger originated FRBs should be estimated
by considering the evolution of both the BNS systems and their magnetic fields.
Based on the population synthesis and adopting a decaying magnetic field of NSs,
we estimate the event rate of BNS-mergers relative to their final magnetic fields.
We find that the rapid merged BNSs tend to merge with high magnetization,
and the event rate of BNS-merger originated FRBs,
i.e., the BNS-mergers with both NSs' magnetic field being higher than $10^{12}\,{\rm Gs}$
is $\sim8\times10^{4}\,\rm{yr}^{-1}$ (19\,\% of the total BNS-mergers) in redshift $z<1$.
\end{abstract}

\section{Introduction}\label{Introduction}
The merger of binary neutron-star (BNS) systems can lead to rich electromagnetic phenomena,
in addition to the strong emission of gravitational waves \citep{Abbott-2017PhRvL, Abbott-2020ApJ}.
In the post-merger phase,
a short gamma-ray burst GRB~170817A \citep{Abbott-2017ApJ, Savchenko-2017ApJ},
the associated afterglows \citep{2017Sci...358.1579H, 2017ApJ...848L..21A, 2019Sci...363..968G, 2018Natur.554..207M},
and a kilonova AT2017gfo peaking at $\sim1\,{\rm day}$ \citep{Abbott-2017ApJ...848L..12A, Coulter-2017Sci}
have been observed as the electromagnetic counterparts of gravitational wave event GW~170817 \citep{Abbott-2017ApJ...848L..12A},
which is the first gravitational wave signal from a BNS merger detected by the advanced LIGO
and Virgo detectors \citep{Abbott-2017PhRvL}.
During the pre-merger phase,
the magnetospheres of two neutron-stars (NSs) in a BNS would interact with each other
and thus energetic Poynting-flux would be driven from the BNS system.
The dissipation of the driven Poynting-flux would appear as multi-band electromagnetic precursors \citep{Hansen-2001MNRAS, 2012ApJ...755...80P}.
This scenario has been confirmed by general relativistic magnetohydrodynamic simulations
(e.g., \citealp{Palenzuela-2013PhRvL, Palenzuela-2013PhRvD, Palenzuela-2014PhRvD, Most-2020ApJ}).
Compared with the post-merger multi-band emission,
radiation signals from the pre-merger BNS may provide more detailed information about
the equation of state for a NS \citep{2022ApJ...939L..25Z, 2022MNRAS.514.5385N, 2022arXiv220808808P} and the magnetospheres interaction in a BNS system \citep{2022arXiv220808808P, 2023MNRAS.519.3923C, 2022arXiv221114433L}.

The interaction of the magnetospheres in a BNS
during its late inspiral phase
could form strong Poynting-flux by extracting the orbit kinetic energy of the system \citep{Palenzuela-2014PhRvD}.
It is shown that the power of the Poynting-flux during this phase
could range from $\sim 10^{38}\,\mathrm{erg\cdot s^{-1}}$ to $10^{44}\,\mathrm{erg\cdot s^{-1}}$ \citep{Palenzuela-2014PhRvD, Wang-2018ApJ},
which is strong enough to form the detectable radio emission from a cosmology distance.
Many efforts have been made for studying the emission from a pre-merger BNS and the association with fast radio bursts (FRBs).
\cite{2013PASJ...65L..12T} proposed that the BNS-mergers with high magnetic field
($B\gtrsim10^{12}\,\rm{Gs}$) could produce the observed FRBs through the coherent radio mechanism like that in a isolated pulsar.
\cite{WangJS-2016ApJ} found that a unipolar inductor model \citep{1969ApJ...156...59G, Lai-Dong-2012,2012ApJ...755...80P}
with a high magnetized primary NS ($B\gtrsim10^{12}\,\rm{Gs}$) and a weak magnetized companion
could produce FRBs by accelerating electrons in coherent slices.
\cite{2023MNRAS.519.3923C} investigated the pulsar-like emission within polar gap model \citep{1975ApJ...196...51R, 1982ApJ...252..337D} and showed the coherent millisecond radio bursts could be detected in Gpc distances by next-generation radio facilities if one NS has a magnetic field higher than $10^{12}\,\rm{Gs}$.
The fast magnetosonic wave has been considered in magnetar \citep{Lyubarsky-2020ApJ, Mahlmann-2022ApJL} and double neutron star system \citep{Most-2022arXiv} as the radio emission mechanism for FRBs.
\cite{Most-2022arXiv} presented the detailed behavior of the flare arising from the magnetic reconnection
in the common magnetospheres of BNS with magnetic fields of $\sim 10^{11}\,\rm{Gs}$,
and demonstrated that the flare interacting with the orbital current sheet
could produce the radio transients with sub-millisecond quasi-periodic structure like FRB 20201020A \citep{2022arXiv220208002P}.

%For the emission formed in a pre-merger BNS,
%the periodicity with a non-repeating trend may be a notable feature
%For the emissions formed in a pre-merger BNS, the periodicity would appear and the event would be one-off
%\citep{Most-2020ApJ, Cherkis-2021ApJ, Lyutikov-2019MNRAS}.
The emissions formed in a pre-merger BNS would appear with periodicity and the corresponding event would be one-off \citep{Most-2020ApJ, Cherkis-2021ApJ, Lyutikov-2019MNRAS}.
Such kind of temporal behavior has been proposed as a possible explanation for the periodicity in the sub-bursts of some one-off FRBs, e.g., FRB~20191221A and FRB~20210213A
\citep{2022Natur.607..256C}.
The BNS-merger originated FRB with periodic sub-pulses may account for a fraction of the FRB population,
in spite of that the estimated rate of BNS mergers ($\sim10^{3}\,\rm{Gpc^{-3}}\,\rm{yr}^{-1}$; \citealp{2022LRR....25....1M, 2019PhRvX...9c1040A}) was significantly less than the rate of FRBs ($\sim10^{4}\,\rm{Gpc^{-3}}\,\rm{yr}^{-1}$; \citealp{2020MNRAS.494..665L}).
In this paper, we investigate the event rate of BNS-merger originated FRBs.
We first study the properties of FRBs~20191221A and FRB~20210213A in the scenario that
a pre-merger BNS is responsible for their radio emission.
The properties of the corresponding BNS is our focus.
Then,
based on the population synthesis and
adopting a decaying magnetic field of NSs,
we estimate the event rates for BNS-mergers relative to their magnetic fields.

The rest of this paper is organized as follows.
In Section~\ref{model},
the luminosity of the Poynting-flux from a pre-merger BNS and the corresponding radio emission
are presented.
Based on the luminosities and periods of two one-off periodic FRBs (FRB~20191221A and FRB~20210213A),
we estimate the magnetic fields of NSs in BNS.
In Section~\ref{distribution}, we estimate the event rate of BNS-merger originated FRBs, based on the population synthesis and adopting a decaying magnetic field of NSs.
In Section~\ref{Discussion}, the conclusion and a brief discussion are presented.

\section{Pre-merger BNS for one-off FRBs}\label{model}

\subsection{Luminosity of the Poynting-flux and the corresponding radio emission}\label{sec:sub2.1}
For a BNS system, the Poynting-flux driven in the pre-merger phase is related to the magnetic field of NSs and the orbital separation of BNS \citep{Palenzuela-2013PhRvL}.
Then, the primary NS with a magnetic dipole moment,
$ \mu _{\ast } = B_{\ast }R_{\ast}^{3} $, and the companion NS with $ \mu _{\mathrm{c} } = B_{\mathrm{c} }R_{\mathrm{c}}^{3} $ are adopted,
where $ B $ and $ R $ are respectively the dipolar magnetic field and radius of a NS,
and $ R_{\ast } = R_{\mathrm{c} } = 13.6\,\mathrm{km}$ is adopted.
In this paper, the subscript ``$\ast$'' (``$ \mathrm{c} $'') represents the parameters of the primary (companion) NS in a BNS, and the primary (companion) NS refers to the NS with a heavy (light) mass in the BNS.

For general BNS systems,
the magnetic field has a negligible effect on the inspiral behavior of BNSs \citep{2000ApJ...537..327I}.
Then, the evolution of the BNS's orbital separation $a$ is mainly associated with the gravitational-wave radiation of the system and thus can be described as (\citealp{1964PhRv..136.1224P}),

\begin{equation}\label{eqn:eqn1}
\dot{a}=-\frac{64 G^{3}( M_{\ast}^{2}M_{\mathrm{c}}+ M_{\ast}M_{\mathrm{c}}^{2})}{5c^{5} a^{3}},
\end{equation}
where $G$ is the gravitational constant,
$ M_{\ast } = M_{\mathrm{c} } = 1.4\,M_{\odot } $ with $M_{\odot}$ being the solar mass is adopted,
and $c$ is the speed of light.
The radius and mass of the NS are obtained with \texttt{LORENE} library\footnote{{LORENE home page, }\href{http://www.lorene.obspm.fr/}{http://www.lorene.obspm.fr/.}}, which assumes a polytropic equation of state $P=K\rho^{\Gamma}$ with $\Gamma=2$ and $K=123$.

Except for the orbital separation,
the Poynting-flux from the system is also dependent on the magnetic field,
dipolar orientations (with respect to the orbital angular momentum), and the ratio of magnetic moments
(\citealp{Palenzuela-2013PhRvL,Palenzuela-2014PhRvD}).
We consider three basic cases in this paper:
$U/u$ case, where the magnetic field in a NS was dominated by that from the other NS, i.e., $ \max \{\mu_{\ast},\mu_{\rm c}\}/a^{3} > \min \{B_{\ast},B_{\rm c}\}$;
$U/U$ case, where the systems with equal magnetic moments aligned with the orbital angular momentum;
and $ U/-U$ case, with an anti-aligned magnetic moment compared to case $ U/U$.
Here, both ``$U$'' and ``$u$'' symbols represent the magnetic dipole moment of a NS,
and ``$U$'' (``$u$'') represents the NS with a high (low) magnetic dipole moment in the BNS if the ``$U$'' appears together with ``$u$''.
In addition, we would like to clarify that the three cases discussed below are particular flavours of the inspiral scenario. Other flavours are not discussed in this paper,
e.g., the pulsar revival model (\citealp{2013PASJ...65L..12T, Hansen-2001MNRAS}).
\begin{itemize}
\item
In the case of $ U/u $,
a unipolar induction model was studied by \cite{Lai-Dong-2012},
in which one of two NSs in the BNS is assumed to be a perfect conductor with a negligible magnetic field.
Following the equation (22) of \cite{Lai-Dong-2012},
the maximum power of the Poynting-flux from the BNS system is given by

\vspace{-0.4cm}
\begin{equation}
L_{\mathrm{BNS} }\approx 4.0 \times 10^{43} \left(\frac{\max \{B_{\ast},B_{\rm c}\}}{10^{12}\ \mathrm{Gs}} \right)^{2}(\frac{a}{27.2 \mathrm{~km}} )^{-7}\ \mathrm{erg}\cdot\mathrm{s}^{-1},\label{eqn:eqn2}
\end{equation}
under the assumption that the resistance of the BNS system is dominated by the magnetospheres,
where $27.2\,\mathrm{km}=2R_{\ast }=2R_{\rm c}$ is the minimum orbital distance for two NSs.
\item
\cite{Wang-2018ApJ} studied the case of $ U/U$.
Based on the equation (5) of \citealp{Wang-2018ApJ}, the power of the Poyting-flux extracted from the system can be read as

\vspace{-0.4cm}
\begin{equation}
L_{\mathrm{BNS}}\approx 1.8 \times 10^{44}(0.19 / \eta-0.08) (\frac{B_{\ast} }{10^{12}\ \mathrm{Gs}  })^{2} (\frac{a}{27.2\mathrm{~km}} )^{-9 / 2}\  \mathrm{erg}\cdot\mathrm{s}^{-1},\label{eqn:eqn3}
\end{equation}
where $ \eta = \Delta r/r$ with $\Delta r$ being the thickness of the compacted region in common magnetosphere
and $r=a/[1+(\mu_{\mathrm{c}}/\mu_{\ast})^{1/3}]=a/2$ represents the distance of both NSs' magnetic fields that come to contact,
and $ \eta = 0.1$ is adopted from \cite{Wang-2018ApJ} which could affect the stored energy in the compacted region.
It should be noted that $\mu_{\rm{c}}=\mu_{\ast}$ and $B_{\rm{c}}=B_{\ast}$ in this case.
%%%%%%%%%%%%%%%%%%%%%%%%%%%%%%%%%%%%%%%%%%%%%%%%%%%%%%%%%%%%%%%%%
\item
For the case $ U/-U$, the twisted magnetic flux loop that connects two NSs can be broken down by the orbit motion. Correspondingly, a flare is ejected by the magnetic reconnection.
The energy stored in the twisted field lines can be estimated as
$ \Delta E_{\text {twist }} \approx  B^{2}_{\ast}R^{3}_{\ast} (2R_{\ast}/a)^{2+\beta } $ \citep{2013ApJ...774...92P}
and the release time-scale of the magnetic field is $ \Delta t\simeq 2a/v_{\mathrm{rec} }$ \citep{2013ApJ...774...92P},
where $ v_{\mathrm{rec} }=0.3c$ is the velocity of reconnection.
Thus, the luminosity of the flare can be estimated as

\vspace{-0.4cm}
\begin{equation}
L_{\mathrm{BNS} }\approx\frac{\Delta E_{\text {twist }}}{\Delta t}=4.1 \times 10^{45} (\frac{a }{27.2\ \mathrm{km}})^{-3-\beta } (\frac{B_{\ast}}{10^{12}\ \mathrm{Gs} } )^{2} (\frac{v_{\mathrm{rec} }}{0.3c} )\ \mathrm{erg}\cdot\mathrm{s}^{-1},\label{eqn:eqn4}
\end{equation}
where
$\beta=1/2$ is the index introduced to match the scaling relation of $ L_{\mathrm{flare} }\propto a^{-7/2}  $ found in \cite{Most-2020ApJ}.
It should be noted that $\mu_{\rm{c}}=\mu_{\ast}$ and $B_{\rm{c}}=B_{\ast}$ in this case.
\end{itemize}
For a more common case $ \mu_{\mathrm{c} } \neq \mu_{\ast  }$,
it is possible to generalized the luminosity estimated in Equations~(\ref{eqn:eqn3}) and (\ref{eqn:eqn4}).
We take the situation with $\mu_{\mathrm{c} } < \mu_{\ast  }$ as an example.
Following the previous works (e.g., \citealp{Palenzuela-2014PhRvD, Wang-2018ApJ}),
a magnetic balance sphere will exist around the companion NS with an effective radius
$a_{\mathrm{eff}} =a/[(\mu_{\ast}/\mu_{\mathrm{c}})^{1/3}+1]$,
and the luminosity from the system with $\mu_{\mathrm{c} } < \mu_{\ast  }$
is nearly equal to that from the system with $ \mu_{\ast  } = \mu_{\mathrm{c} }$ and $a=2a_{\mathrm{eff}} $.
That is to say, Equations~(\ref{eqn:eqn3}) and (\ref{eqn:eqn4}) can be used in a system with unequal magnetic fields
by replacing $B_{\ast  }$ with $B_{\mathrm{c} }$ and $a$ with $2a_{\mathrm{eff}}$.
It should be noted that the luminosity of Poynting-flux
is subject to the minimum magnetic moment of the NS (i.e., $\min\{\mu_{\ast  },\mu_{\mathrm{c} }\}$)
for both the case $ U/-U$ and $U/U$,
and to the maximum magnetic moment of the NS (i.e., $\max\{\mu_{\ast  },\mu_{\mathrm{c} }\}$) for the case $ U/u$.
In this paper,
we consider the BNS with equal magnetic fields of NSs,
which should be treated as the lower limit of the magnetic fields in the BNS system.
In addition, the effect of the NS spin on the Poynting-flux from the BNS system (\citealp{Most-2020ApJ, Cherkis-2021ApJ, 2022MNRAS.515.2710M}) is not considered in this paper.

For FRBs originated from pre-merger BNSs,
the radio emission is formed during the dissipation of the Poynting-flux
driven during the inspiral phase of a BNS-merger.
We assume that the fast magnetosonic wave is responsible for the radio emission,
which was used to explain the giant radio pulse from pulsar
(\citealp{2019ApJ...876L...6P, 2019MNRAS.483.1731L, 2003Natur.422..141H}).
In the spirit of \cite{2019ApJ...876L...6P}, we take the following relation
between FRB luminosity $L_{\mathrm{FRB} }$ and Poynting-flux luminosity $L_{\mathrm{BNS} }$, i.e.,

\begin{equation}
L_{\mathrm{FRB} }=f\frac{v_{\mathrm{rec}}}{c}\frac{L_{\mathrm{BNS}}}{1/(4\Gamma^{2}_{\mathrm{BNS}})} =2.4\times10^{-3} \frac{f}{0.002} \frac{v _{\mathrm{rec} }}{0.3c}\Gamma^{2}_{\mathrm{BNS} } L_{\mathrm{BNS} },\label{eqn:eqn6}
\end{equation}
where $f$ is the efficiency of the magnetic reconnection driving the fast magnetosonic wave
and $\Gamma_{\mathrm{BNS}} $ is the bulk Lorentz factor of the Poynting-flux.
According to \cite{Lyubarsky-2020ApJ}, an outward Poynting-flux,
which arrived at the light cylinder $R_{\mathrm{LC} } = \mathrm{c}/\Omega_{\mathrm{orbit}} $ with magnetic field strength $B_{\rm BNS}=\sqrt{L_{\rm BNS}/c }/R_{\rm LC }$,
could have a bulk Lorentz factor $\Gamma _{\mathrm{BNS} }=\mathrm{max} [\sqrt{{B_{\mathrm{BNS}}}/{B_{\mathrm{LC} }} }/2 ,\Gamma _{\mathrm{BNS,min} }]$,
where $\Omega_{\mathrm{orbit} } =[{G M_{\ast}(1+M_{\mathrm{c} }/M_{\ast})}/{a^{3} } ]^{{1}/{2} }$ is the orbital frequency and $ \Gamma _{\mathrm{BNS,min} }=2 $ is the minimum Lorentz factor.
Once the outflow moves across the light cylinder,
the perturbation in the current sheet could trigger the magnetic reconnection,
and the following fast magnetosonic wave will escape into the vacuum and convert to the coherent emission \citep{Most-2022arXiv, 2019ApJ...876L...6P, 2019MNRAS.483.1731L}.
In this process,
local kinetic simulation \citep{Mahlmann-2022ApJL} showed that the conversion efficiency of the reconnection energy to the fast magnetosonic wave energy is $f \simeq 0.002$.

\subsection{Properties of BNS for two one-off FRBs}\label{sec:sub2.2}
There are three one-off FRBs with periodic sub-pulses, i.e., FRBs 20191221A ($216.8~\rm{ms}$), 20210206A ($2.8~\rm{ms}$), and 20210213A ($10.7~\rm{ms}$),
reported by \cite{2022Natur.607..256C},
where the period signal significance of FRB~20191221A ($6.5\,\sigma$) is higher than that of FRB~20210206A ($1.3~\sigma$) and FRB~20210213A ($2.4~\sigma$).
It is an open question for the origin of their periodicity
\citep{2022arXiv221107669B, 2023MNRAS.519.3923C, 2022arXiv221009930W}.
It is proposed that such kinds of FRBs (like FRB~20191221A and FRB~20210213A) may be produced during the inspiral phase of a BNS-merger (\citealp{2022Natur.607..256C, 2022MNRAS.515.2710M}), and we noted that FRB 20210206A is disfavoured in a BNS-merger scenario as its expected period derivative is not consistent with the observed sub-pulse separation (\citealp{2022Natur.607..256C}).
In this scenario,
we study the properties of the corresponding BNS system based on these two FRBs' periodicity and luminosity (FRB~20191221A and FRB~20210213A)\footnote{ FRB 20201020A is detected with sub-millisecond period ($\sim0.415\,\rm{ms}$) and
high period signal significance ($2.5\,\sigma$) \citep{2022arXiv220208002P}.
However, this burst is not involved in this section since its period
may be too small compared with the minimum orbital period of BNSs (\citealp{2020ARNPS..70...95R}).}.

\emph{Periodicity Estimation.\,}
The direction of the Poynting-flux is dependent on the relative inclination of the magnetic moments and their ratio \citep{Palenzuela-2014PhRvD, Most-2020ApJ}.
In general, the magnetic dipole axis is not aligned with the spin axis in pulsar magnetic field configuration,
this suggests that the timescale of Poynting-flux powering flares in the light of sight should be closed to the orbit period $P_{\rm{orbit}}$ of the corresponding BNS system \citep{Palenzuela-2014PhRvD}.
For simplicity, we assume that the period $P_{\rm{FRB}}$ of these two FRBs were corresponded to the orbit period of the corresponding BNS system $P_{\rm{orbit}}={2\pi }{[{G(M_{\ast}+M_{\rm{c}})}/{a^{3} } ]^{-\frac{1}{2} }}$, i.e., $P_{\rm{FRB}}=P_{\rm{orbit}}$.

\emph{Luminosity.\,}
The isotropic peak luminosity of these bursts is estimated as follows \citep{2018ApJ...867L..21Z},

\begin{equation}
   L_{\mathrm{FRB} } \simeq  4\pi \times 10^{42} (\frac{D_{\mathrm{L}} }{10^{28}\ \mathrm{cm}})^{2 } \frac{F_{\mathrm{p}} }{\mathrm{Jy} } \frac{\nu_{\mathrm{cf} }}{\mathrm{GHz} } \ \mathrm{erg}\cdot\mathrm{s}^{-1},\label{eqn:eqn5}
\end{equation}
where $F_{\mathrm{p}} $ is the peak flux, $ D_{\mathrm{L}} $ is the luminosity distance of the burst,
and $ \nu_{\mathrm{cf}} $ is the central frequency in observed bandwidth.
The value of $ D_{\mathrm{L}} $ is estimated with the redshift $z$,
and the redshift of the FRBs with host-galaxy location is found to be related to
the excess dispersion measure $\rm DM_{\rm excess}$, i.e., the total dispersion measure without the contribution from our Galaxy \citep{2020Natur.581..391M}.
The relation between $z$ and $\rm{DM}_{\rm{excess}}$ obtained from the appendix A of \cite{2022Ap&SS.367...66C},
i.e., ${\rm DM}_{\rm{excess}}=1028z+84.34$, is used in this paper,
where the contributions of our Galaxy on the dispersion measure for these two FRBs is from \cite{2017ApJ...835...29Y}\footnote{\href{https://www.atnf.csiro.au/research/pulsar/ymw16/}{https://www.atnf.csiro.au/research/pulsar/ymw16/}}.
Correspondingly, the obtained redshifts of FRB~20191221A and FRB~20210213A are $ \sim 0.19 $ and $ \sim 0.35 $, respectively.
Since the maximum redshift of FRB is $\sim 1.0$ (FRB20220610A, \citealp{2022arXiv221004680R}), the BNS-mergers within a redshift of 1 is our focus.

\begin{figure*}
\centering{
\begin{tabular}{ccc}
\includegraphics[width=.32\linewidth]{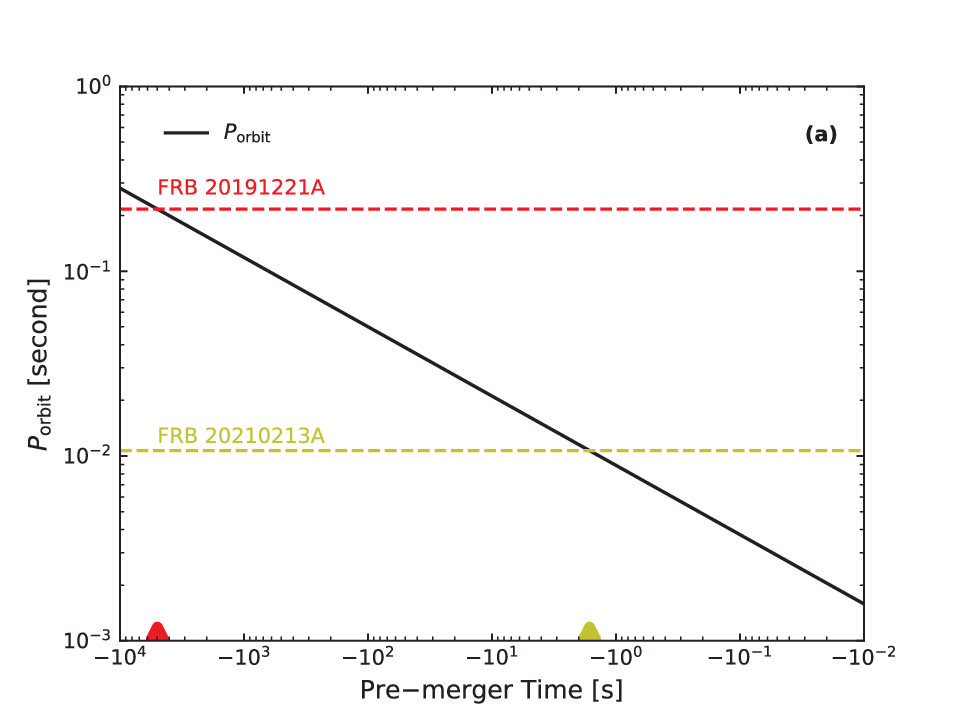} &
\includegraphics[width=.32\linewidth]{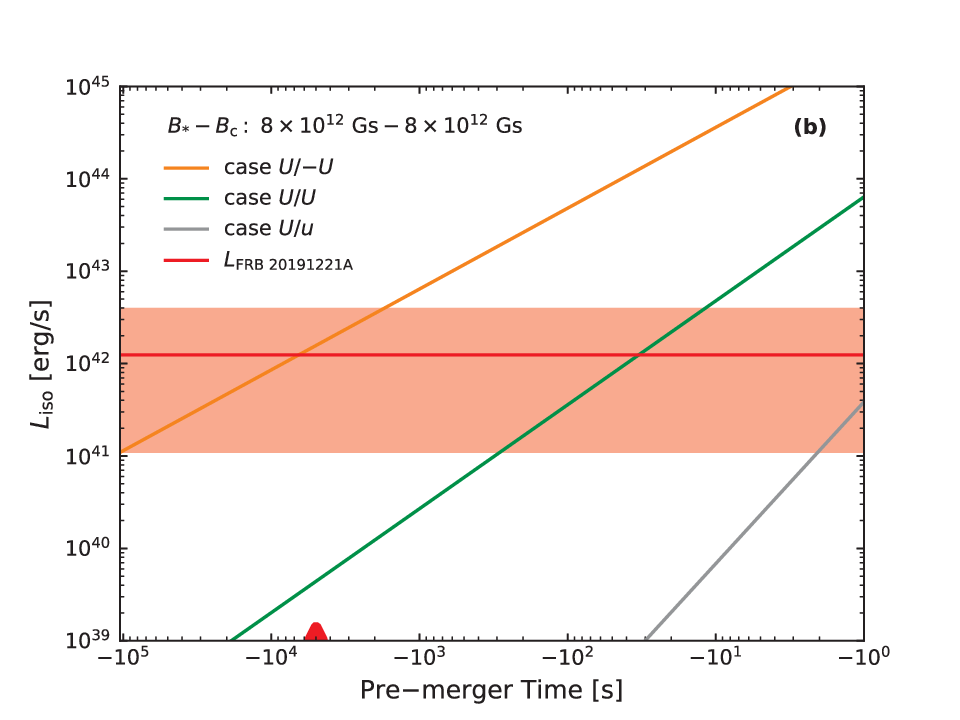} &
\includegraphics[width=.32\linewidth]{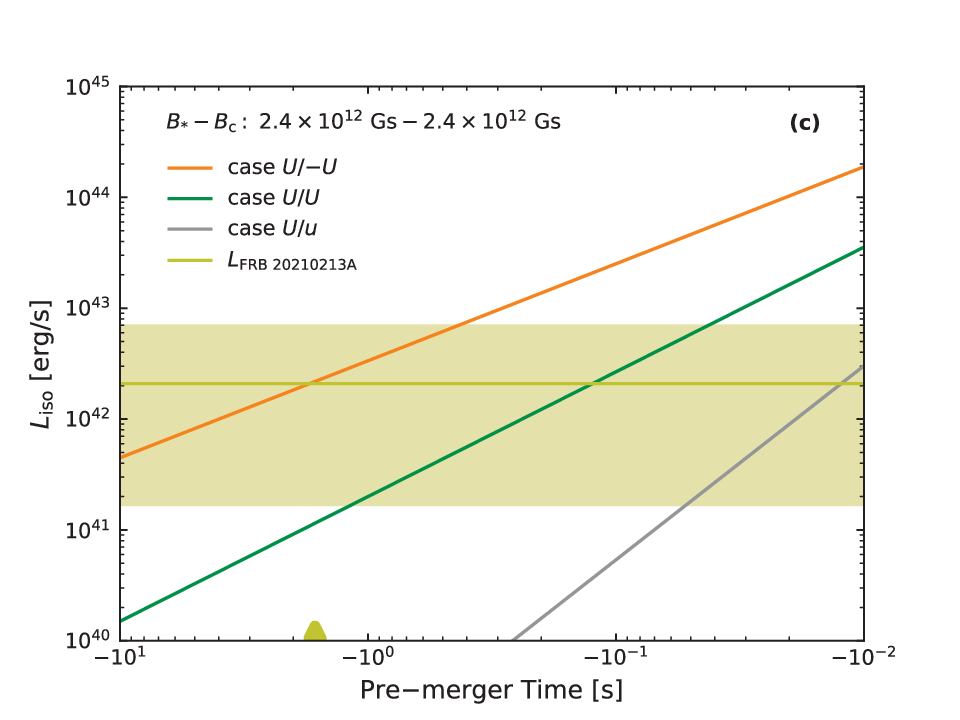}
\end{tabular}}
\caption{Panel-(a): the relation of orbit period (or one-off FRBs' period) and the pre-merger time (black line),
where the periods of FRB~20191221A and FRB~20210213A are plotted with horizontal dash lines
in red, blue, and yellow color, respectively.
The solid triangles in the bottom of panel mark the appearance times at which $P_{\rm{orbit}}=P_{\rm{FRB}}$ for these FRBs,
which is also used in panels-(b) and (c).
In panels-(b) and (c), we show the minimum magnetic fields of our studied FRBs
in different cases,
where the horizontal lines and the corresponding shaded region
show the isotropic peak luminosity of these FRBs and
that with $68\%$ uncertain due to $\rm{DM}_{excess}$-redshift relation (see Figure 5 in \citealp{2022Ap&SS.367...66C}), respectively.
Since the minimum magnetic field is required in the case of $U/-U$ (orange lines),
we used this case to match the peak luminosities of our studied FRBs in their appearance times and obtained the corresponding magnetic fields (shown in the left-upper part of each panel).
%With the same magnetic fields, the luminosity relative to the pre-merger time for the case of $U/U$ (green lines) and the case of $U/u$ (gray lines) are also shown in each panel.
%\cite{2022ApJ...939L..25Z} proposed that the precursor of GRB~211211A
%should be due to the tidal-induced NS's crust destruction at $\sim0.1-0.3$~s before the coalescence,
%a possible moment for forming the precursor of GRB 211211A is also indicated with a vertical magenta line at $-10^{-1}$~s.
}\label{MyFigA}
\end{figure*}

\emph{Properties of BNSs for FRBs.\,}
The luminosity of a pre-merger BNS would be related to the value of $a$ and thus the appearance time of FRBs preceding the BNS-mergers.
Then, we first estimate the appearance times of our studied FRBs
preceding the BNS-mergers based on the periods of these FRBs.
In the panel-(a) of Figure~{\MyFigA},
we plot the relation of the periods of FRBs and the orbital period of the BNS with black line.
With the periods of our studied FRBs,
the orbital period of the BNS corresponding to FRB is estimated.
From this panel, one can find that the appearance times of FRBs~20191221A (red triangle) and 20210213A (blue triangle) are $\sim {-5000}\,{\rm s}$ and $-1.6\,{\rm s}$, respectively.
Here, the merger time of BNS is set as zero time.
In panels-(b) and (c) of Figure~{\MyFigA}, we show the minimum magnetic fields of NSs in the BNS for our studied FRBs.
In these panels, the luminosity of Poynting-flux relative to the pre-merger time of BNS
in the case of $ U/u $, $ U/U $, and $U/-U$
with a same set of $B_{*}$ and $B_{\rm c}$
is shown with gray, red, and orange lines, respectively.
From these lines,
the luminosity of Poynting-flux in the case of $U/-U$ (orange lines)
is higher than those in other two cases.
Thus, the minimum strength of magnetic fields of NSs
is estimated based on the case of $U/-U$
and shown in the upper-left part of these two panels,
i.e., $8\times10^{12}\,\rm{Gs}$ and $2.4\times10^{12}\,\rm{Gs}$
for FRB~20191221A and FRB~20210213A, respectively.
Then, we can conclude that the minimum strength of the magnetic field of NSs in the BNS
that is responsible for these two one-off FRBs
should be around $\gtrsim10^{12}\,\rm{Gs}$.
Such high magnetic field required for explaining one-off FRBs are also proposed in a serial of works,
e.g., \cite{2013PASJ...65L..12T} and \cite{Lyutikov-2019MNRAS}, and \cite{Cherkis-2021ApJ}.
We also found that it is difficult for case $U/u$ to produce FRBs like FRB~20191221A and FRB~20210213A
since a magnetar with $\gtrsim10^{14-16}\,\rm{Gs}$ should be involved in the corresponding BNS-merger.
Suck kind of BNS-mergers would be just a tiny fraction ($\sim0.35\,\%$) of total BNS-mergers at $z<1$ based on the estimation in Section~\ref{distribution} and thus is not the focus in Section~\ref{distribution}.
Hereafter, ``the BNS-merger originated FRBs'' refers to the FRBs formed in the case $U/-U$ or $U/U$.

%It should be noted that the sub-pulse seperation of FRB~20210206A is different from that of FRB~20191221A and FRB~20210213A.
%\cite{2022Natur.607..256C} has fit the arrival times of the sub-pulses for these three FRBs,
%and found that FRB 20191221A and FRB 20210213A are well-fitted by a BNS-merger scenario,
%while FRB 20210206A is disfavoured
%due to its expected period derivative is not consistent with the observed sub-pulse separation.
%It is interesting to point out that
%the appearance time of FRB~20210206A is located within the trigger time of GRB~211211A's precursor (see Figure~{\MyFigA}),
%which is suggested to be formed during a catastrophic global crust destruction on the NSs
%preceding the BNS-merger for GRB~211211A (\citealp{2022ApJ...939L..25Z}).
%It means that FRB 20210206A maybe come from a very closed BNS system.
%For a very closed BNS system,
%the strong tidal force could break the NS's crust and
%thus deform the NS's magnetosphere \citep{2022arXiv220502186X, 2022ApJ...939L..25Z, 2022A&A...664A.177S, 2012PhRvL.108a1102T},
%which results in the change of the radio mechanism and its period.
%It could be a possible explanation for the sub-pulse separation of FRB 20210206A.

We have found that the high magnetic field ($\gtrsim10^{12}\,\rm{Gs}$) in a pre-merger BNS is needed for producing the one-off FRBs we studied.
Such a high magnetic field is often observed in the Galactic isolated pulsar, while it is relatively rare for the pulsar in the Galactic BNS-systems (\citealp{ATNF-2005AJ}).
This implies that most of the BNSs in our Galaxy would merge with low magnetic fields\footnote{\href{https://www.atnf.csiro.au/research/pulsar/psrcat/}{https://www.atnf.csiro.au/research/pulsar/psrcat/}} ($\sim10^{9}\,\rm{Gs}$; \citealp{ATNF-2005AJ})
and could not power the FRBs we studied.
To obtain a credible event rate of BNS-merger originated FRBs, i.e., BNS-mergers with both NSs' magnetic field strength being higher than $10^{12}\,\rm{Gs}$, we use the population synthesis and adopting a decaying magnetic field of NSs to study the event rate of BNS-mergers in Section~\ref{distribution}.
We also provide the event rate of BNS-mergers with at least one magnetized NS ($>10^{12}\,\rm{Gs}$),
which could be useful in other scenarios, e.g., \cite{WangJS-2016ApJ, 2023MNRAS.519.3923C}.

\section{Merger Rate of BNSs relative to their Magnetic field}\label{distribution}
\emph{Magnetic field Prescription for a NS.\,}
In order to estimate the event rate of BNS-merger originated FRBs,
the event rate of BNS-mergers relative to the final magnetic fields of the NSs should be given.
% We use the data (Model M33.A) of the BNS-mergers given in the population synthesis of \cite{2020A&A...636A.104B},
% which is consistent with the observations of LIGO-Virgo O1/O2 observing runs \citep{2019PhRvX...9c1040A}.
We use the data of the BNS-mergers from Model M33.A in \cite{2020A&A...636A.104B},
of which the main features of the binary evolution models for Model M33.A can be found in their table~2
and the binary evolution calculations were performed with the upgraded population synthesis code \texttt{StarTrack}\footnote{The \texttt{StarTrack} population synthesis code is not an open source code and the basic description of the code can be found in \cite{2002ApJ...572..407B} and \cite{2008ApJS..174..223B}. The improvements of \texttt{StarTrack} was given in the Appendix A.7 of \cite{2020A&A...636A.104B}.
The \texttt{StarTrack} population synthesis code was developed for the study of double compact object mergers
based on binary evolution calculations.
Given models about the cosmic star formation history and metallicity evolution,
the simulation of \texttt{StarTrack} tracks the mergers of double compact objects in different redshift.
}
\citep{2002ApJ...572..407B, 2008ApJS..174..223B}.
Based on the date of Model M33.A, the BNS-mergers rate density ($\rm{Gpc^{-3}\,yr^{-1}}$) in different redshift
can be obtained and is plotted in the right-panel of Figure~{\MyFigB} with black line.
The local merger rate densities of BNS-mergers and binary black hole mergers from Model M33.A (\cite{2020A&A...636A.104B}) are consistent with the observational limits of LIGO-Virgo O1/O2/O3 observing runs (\citealp{2019PhRvX...9c1040A, 2021arXiv211103634T, 2022LRR....25....1M}).
Besides, the magnetic field of the NSs in the BNS-mergers is vital in our estimation.
Then, the magnetic field of a newborn NS and its evolution during the lifetime of the corresponding NS should be prescribed.
\cite{2011MNRAS.413..461O} modeled the population of BNSs in our Galaxy by
using the binary population synthesis code \texttt{StarTrack} \citep{2002ApJ...572..407B, 2008ApJS..174..223B},
and found that the scenario with an exponentially decaying magnetic field for NSs is consistent with the observation.
Then, the magnetic field evolution of NSs is took as

\begin{equation}
B_{\ast/\mathrm{c} }=\left ( B_{\mathrm{init} }-B_{\mathrm{min} } \right ) \exp \left ( \frac{t_{\ast/\mathrm{c} }}{\Delta }  \right ) +B_{\mathrm{min}},\label{eqn:eqn7}
\end{equation}
where $ B_{\mathrm{init}} $ is the magnetic field of the newborn NS,
$ \Delta $ is the decay timescale of the magnetic field (\citealp{1986ApJ...305..235T, 1997MNRAS.292..167U, 1997MNRAS.284..311K, 1999MNRAS.308..795K, 2002ApJ...565..482G, 2004ApJ...604..775G, 2018MNRAS.481.4009V}; see \citealp{2021Univ....7..351I} for a review),
and $ t_{\ast} $ ($ t_{\mathrm{c} } $) is the lifetime of the primary NS (the companion NS) in BNSs.
The value of $ B_{\mathrm{init}}/\rm Gs$ is randomly selected in a lognormal distribution
with mean $\mu_{B_{\rm init}}= 12.66 $ and variance $\sigma_{B_{\rm init}}= 0.35 $;
the value of $ \Delta/\rm{Myr}$ is also randomly selected from a lognormal distribution
with mean $\mu_{\Delta}=0.56 $ and variance $\sigma _{\Delta}= 0.075 $\footnote{The value of $\log\left ( \sigma _{\Delta}  \right /\mathrm{Myr} ) = 0.075 $
is estimated by fitting the distribution of $ \log\left ( \Delta  \right /\mathrm{Myr} )$ in the ``Power-law model''
of figure~4 in \cite{2020MNRAS.492.4043C}.} (\citealp{2020MNRAS.492.4043C}).
It should be noted that
the lognormal distribution of the initial magnetic field and the corresponding decay timescale are obtained based on the isolated pulsars
from simulations (\citealp{2020MNRAS.492.4043C}) and observations (\citealp{2022MNRAS.514.4606I}).
The observation of Galactic NSs reveals that the isolated NSs and the NSs from BNS present an almost same behavior in the relation of the magnetic field and NS's characteristic age\footnote{\href{https://www.atnf.csiro.au/research/pulsar/psrcat/}{https://www.atnf.csiro.au/research/pulsar/psrcat/}} \citep{ATNF-2005AJ}.
We set a maximum value $ B_{\mathrm{init,max}}=10^{14}\,\mathrm{Gs}$ (i.e., $ B_{\mathrm{init}}\leqslant B_{\mathrm{init,max}}$) for the initial magnetic field of the newborn NS,
where the $10^{14}\,\mathrm{Gs} $ is the maximum value we used according to the high-B pulsars (a rough definition is given in Section 3.4 of \citealp{2019RPPh...82j6901E}).
The minimum magnetic field $B_{\mathrm{min}}$ is drawn from a logarithmic uniform distribution between $ 10^{7}\,\mathrm{Gs} $ and $ 10^{8}\,\mathrm{Gs} $ \citep{2006MNRAS.366..137Z,1994MNRAS.269..455J}.
The lifetime of the primary NS (the companion NS) in BNSs, i.e., $ t_{\ast} $ ($ t_{\mathrm{c} } $), was obtained based on the data of Model M33.A in \cite{2020A&A...636A.104B},
of which both the born-time and the merger-time of a NS in every BNS-merger are recorded.
It should be noted that the impact of the NS's magnetic field on the binary evolution is not involved in the {\tt StarTrack} population synthesis code and thus in this work.
We noted that the BNS-mergers rates below were calculated in the local universe, i.e., $z=0$, and the calculation was based on the section 2.2 of \citep{2016ApJ...819..108B}.

\begin{figure*}
\centering{
\begin{tabular}{ccc}
\includegraphics[trim=20 0 0 0, scale=0.4]{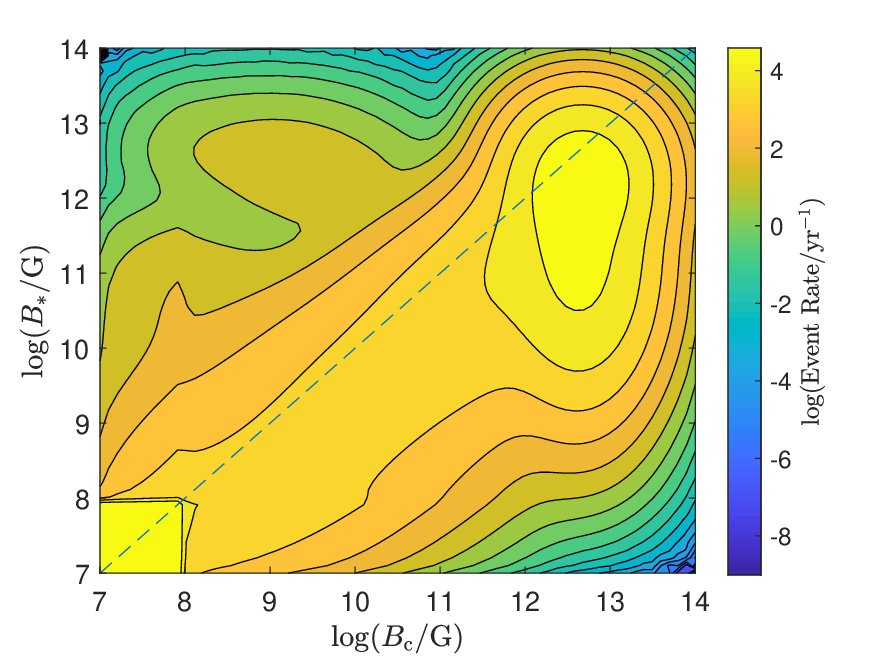} &
\includegraphics[trim=20 0 0 0, scale=0.4]{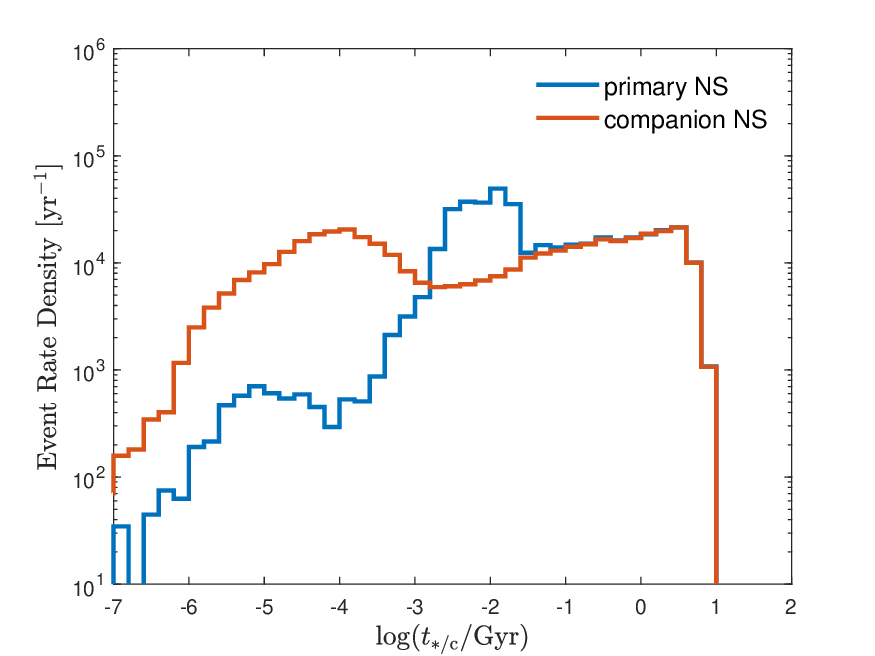} &
\includegraphics[trim=20 0 0 0, scale=0.4]{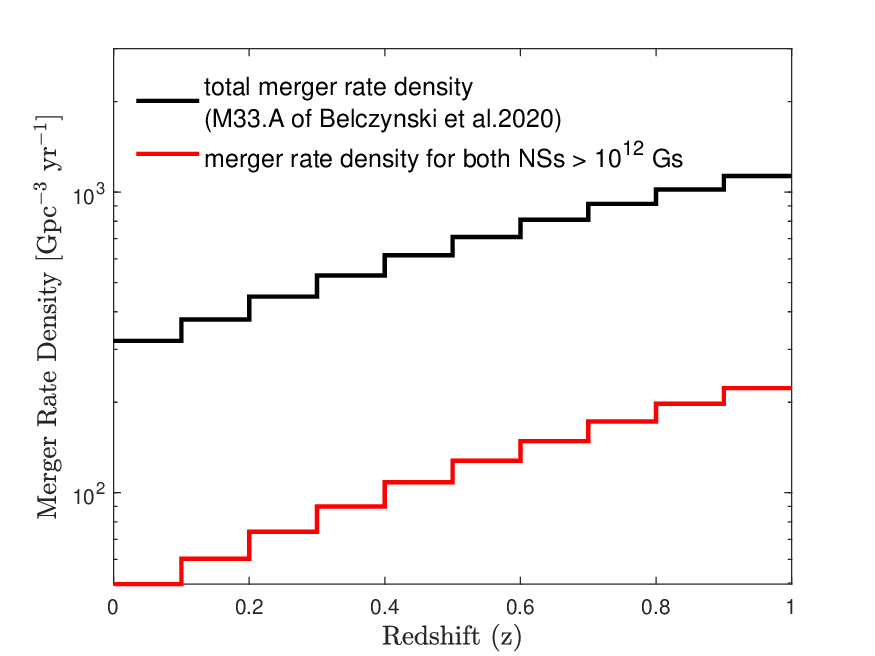} \\
\end{tabular}}
\caption{\emph{Left-panel}--- the event rate density of BNSs (color bar) with different MF strength of the primary star and the companion at redshift $z<1$,
where the dashed line represents the relation of $B _{\ast}=B _{\mathrm{c}}$,
and the event rate of BNSs for $B_{\ast}\in [B_{\ast,l}, B_{\ast,u}]$
and $B_{\rm c}\in[B_{{\rm c},l}, B_{{\rm c},u}]$ can be obtained by multiplying the event rate density
with $\log(B_{{\rm c},u}/B_{{\rm c},l})\times \log(B_{\ast,u}/B_{\ast,l})$.
Noting that the regular square region in the left-bottom of this panel
is formed due to that a flat logarithmic distribution of $ B_{\mathrm{min}}$ is adopted in our calculations.
\emph{Middle-panel}--- the event rate density distributions relative to the lifetime of the primary NS (blue line) and companion NS (orange line) in the BNS-mergers.
It shows that the lifetime distribution of the companion NS in the BNS-mergers has two peaks at $t_{\rm c}\sim 10^{-4}\,{\rm Gyr}$ and $10^{0.5}\,{\rm Gyr}$, while that of the primary NS has three peaks at $t_{*}\sim 10^{-5}\,{\rm Gyr}$, $t_{*}\sim 10^{-2}\,{\rm Gyr}$, and $10^{0.5}\,{\rm Gyr}$.
\emph{Right-panel}--- the BNS-mergers rate density ($\rm{Gpc^{-3}\,yr^{-1}}$) relative to the redshift at $z<1$. The total BNS-mergers rate density (black line) is plotted based on the data (Model M33.A) in \cite{2020A&A...636A.104B} and our estimated BNS-mergers rate density with both NSs' magnetic fields being larger than $10^{12}\,\rm{Gs}$ is shown with red line.
}
\label{MyFigB}
\end{figure*}

\emph{Event Rate of BNS-mergers relative to their final magnetic fields.\,}
In the left panel of Figure~{\MyFigB},
we plot the event rate density of BNS-mergers relative to the magnetic field strength of the primary NS ($B_{*}$) and that of the companion NS ($B_{\rm c}$) within the redshift of ${1.0}$.
Here, the dash line represents the relation of $B _{\ast} = B _{\mathrm{c} }$,
and the event rate of BNS-mergers for $B_{\ast}\in [B_{\ast,l}, B_{\ast,u}]$
and $B_{\rm c}\in[B_{{\rm c},l}, B_{{\rm c},u}]$ can be obtained by multiplying the event rate density
with $\log(B_{{\rm c},u}/B_{{\rm c},l})\times \log(B_{\ast,u}/B_{*,l})$.
Based on the result shown in this panel, one can find the following facts.
(1) The BNS-mergers concentrate in the field of $(B_{*}, B_{\mathrm{c}})\sim (10^{12}\,\mathrm{Gs}, 4\times 10^{12}\,\mathrm{Gs})$.
The event rate of BNS-mergers in this field, i.e., $\log\left ( B_{\ast}  \right /\rm{Gs} )\in12\times[0.32,1.68]$ and $\log\left ( B _{\ast}  \right /\rm{Gs} )\in12.6\times[0.32,1.68]$,
is $ \simeq 4.6\times10^{4}\,\mathrm{yr}^{-1}$ and $11\,\%$ of the total BNS-mergers.
(2) The event rates of the BNS-mergers with both $B_{*}$ and $B_{\rm c}$
being higher than $ 10^{11}\,\mathrm{Gs} $, $ 10^{12}\,\mathrm{Gs} $, and $ 10^{13}\,\mathrm{Gs} $
can be estimated,
and are $ \sim1.6\times10^{5}\,\mathrm{yr}^{-1}$ ($40\,\%$ of the total BNS-mergers),
$ \sim 8\times10^{4}\,\mathrm{yr}^{-1}$ ($19\,\%$ of the total BNS-mergers),
and $\sim 567\,\mathrm{yr}^{-1}$ ($0.1\,\%$ of the total BNS-mergers), respectively.
(3) The event rates of the BNS-mergers with either $B_{*}$ or $B_{\rm c}$
being higher than $ 10^{11}\,\mathrm{Gs} $, $ 10^{12}\,\mathrm{Gs} $, and $ 10^{13}\,\mathrm{Gs} $
are $ \sim 2.2\times10^{5}\,\mathrm{yr}^{-1}$ ($52\,\%$ of the total BNS-mergers),
$ \sim 2.0\times10^{5}\,\mathrm{yr}^{-1}$ ($47\,\%$ of the total BNS-mergers),
and $ \sim 3.1\times10^{4}\,\mathrm{yr}^{-1}$ ($7\,\%$ of the total BNS-mergers), respectively.
We also calculate the event rate within the maximum redshift of our studied FRBs, i.e., $z\leqslant0.35$.
It is found that the fraction of BNS-mergers related to different final magnetic fields of NSs is nearly consistent with the result estimated in the above.

In the left panel of Figure~{\MyFigB}, more than half of BNS-mergers with both $B_{*}$ and $B_{\rm c}$
being higher than $10^{11}\,\rm{Gs}$ is distributed below the dash line, i.e., $B _{*}=B _{\mathrm{c} }$.
It reveals that the companion NS tend to own a higher magnetic field strength compared with that of the primary NS.
This is no surprise because the companion NS in a BNS system is born behind the primary NS and thus
the decay of its magnetic field is relatively weak compared with that of the primary NS.
In the middle panel of Figure~{\MyFigB}, we show the event rate density distributions relative to
the lifetime of the primary NS (blue line) and the companion NS (orange line) in BNS-mergers.
The event rate for a specific lifetime interval (with a unit of Gyr) could be obtained by multiplying the event rate density with its logarithmic dimensionless time interval.
It is shown that the lifetime distribution of the companion NS in the BNS-mergers has two peaks,
i.e., peak at $t_{\rm c}\sim 10^{-4}\,{\rm Gyr}$ with event rate density of $\sim 2\times10^{4}\,\rm{yr}^{-1}$
and peak at $t_{\rm c}\sim 10^{0.5}\,{\rm Gyr}$ with event rate density of $\sim 2\times10^{4}\,\rm{yr}^{-1}$.
The lifetime distribution of the primary NS in the BNS-mergers has three peaks,
i.e., peaking at $t_{*}\sim 10^{-5}\,{\rm Gyr}$ with event rate density of $\sim5\times10^{2}\,\rm{yr}^{-1}$,
at $t_{*}\sim 10^{-2}\,{\rm Gyr}$ with event rate density of $\sim5\times10^{4}\,\rm{yr}^{-1}$,
and at $t_{*}\sim 10^{0.5}\,{\rm Gyr}$ with event rate density of $\sim2\times10^{4}\,\rm{yr}^{-1}$.
It reveals that the BNS-mergers can be divide into two kinds of mergers based on the lifetime of the companion NS:
the rapid merged BNSs and the slow merged BNSs.
In the rapid merged BNSs, the lifetime of the companion NS peaks at $t_{\rm c}\sim 10^{-4}\,{\rm Gyr}$.
Since the lifetime of the companion NS is less than the decay timescale of a NS (i.e., $\Delta\sim 10^{-2.4}\,{\rm Gyr}$),
the magnetic field of the companion NS in the rapid merged BNSs is almost the same as that at its birth, i.e., the initial magnetic field.
However, the magnetic field of the primary NS is generally suffered from significantly decaying in two kinds of mergers.
Then, the following two facts can be understood.
(1) The rapid merged BNSs are always presented with a strong magnetized companion NS and a weak magnetized primary NS. Based on the middle panel of Figure~{\MyFigB},
one can find that the rapid merged BNSs is 47\,\% of the total BNS-mergers.
Then, one can expect that the BNS-mergers with one NS's final magnetic field being higher than $10^{12}\,\rm{Gs}$ should be around 50\,\% of the total BNS-mergers. This is indeed obtained.
(2) The BNS-mergers with both NSs' magnetic field being higher than $10^{12}\,\rm{Gs}$ should be related to
the rapid merged BNSs with both NSs' lifetime being closed to the decay timescale of a NS
(i.e., $\Delta\sim 10^{-2.4}\,{\rm Gyr}$).
Based on the middle panel of Figure~{\MyFigB},
one can find that such kind of the rapid merged BNSs is around $19\,\%$ of the total BNS-mergers.
In the right panel of Figure~{\MyFigB}, we plot the total BNS-mergers rate density (black line) and the BNS-mergers rate density with both NSs' magnetic fields being
larger than $10^{12}\,\rm{Gs}$ (red line), both of which are related to a redshift at $z<1$.

Magnetar was not considered in Figure~{\MyFigB}, because the exact condition for a progenitor to produce the magnetic field in magnetar (or pulsar) is still unclear \citep{2021Univ....7..351I, 2022ApJ...926..111W, 2021MNRAS.504.5813M, 2005ApJ...620L..95G, 2006ApJ...636L..41M}.
Here, we show that involving magnetars in our estimation of the even rate of BNS-mergers relative to the final magnetic field could not obviously change our results presented above.
Magnetar is a strong magnetized ($10^{14}{-}10^{15}\,\mathrm{Gs} $) NS with period of several seconds \citep{Magnetars-2017ARA&A}, which may have an unusually evolutionary behavior in its magnetic field compared with that of the pulsar.
It is found that
the decay timescale of a magnetar ($\Delta_{\rm{mag}}=4\,\rm{kyr}$; \citealp{2022MNRAS.509..634J}) is shorter than
that of a pulsar ($\Delta\simeq4\,\mathrm{Myr}$; \citealp{2020MNRAS.492.4043C}) with around three orders of magnitude.
Although the companion of a magnetar has not been found\footnote{\href{https://www.physics.mcgill.ca/~pulsar/magnetar/main.html}{https://www.physics.mcgill.ca/~pulsar/magnetar/main.html}} \citep{2014ApJS..212....6O, 2022MNRAS.513.3550C},
a magnetar is still possibly born in BNS-systems through core-collapse supernovae \citep{2015PASJ...67....9N, 2019MNRAS.487.1426B} or the accretion collapse of ONe white dwarfs \citep{Duncan-1992ApJ, 2006MNRAS.368L...1L, 2022MNRAS.509.6061A}.
We assume that the magnetar can born in the BNS-systems and estimate the event rate of rapid merged BNS with a magnetar ($\gtrsim10^{14}\,\rm{Gs}$),
which should be the BNSs with the merger time being $\lesssim \Delta_{\rm{mag}}=4\,{\rm kyr}$.
This is owing to that the magnetic field of a magnetar would decay significantly
in the BNSs with the merger time being larger than $\Delta_{\rm{mag}}=4\,{\rm kyr}$.
The event rate of BNS-mergers with the merger time being $\lesssim \Delta_{\rm{mag}}=4\,{\rm kyr}$ is
estimated to be ${\cal R}_{\rm mag}\approx  1.4\times 10^{3}(f_{\rm mag}/0.1)\,\rm yr^{-1}$ or 0.35\,\% of the total BNS-mergers,
where $f_{\rm{mag}}=0.1$ is
the adopted as the minimum fraction of magnetar in the young NS population \citep{Magnetars-2017ARA&A}.
Compared with the rapid merged BNSs,
the fraction of the BNS-mergers with the merger time being $\lesssim \Delta_{\rm{mag}}=4\,{\rm kyr}$ (i.e., the BNS-mergers with a magnetar) can be neglected.

\section{Discussions and Conclusions}\label{Discussion}
It is proposed that one-off FRBs with periodic structures
may be produced during the inspiral phase of a BNS-merger.
In this paper, we study the event rate of such kind of FRB.
We first investigate the properties of some one-off FRBs with periodic sub-pulses
(i.e., FRB~20191221A and FRB~20210213A)
based on the scenario that the Poynting-flux from pre-merger BNSs drives the observed radio emission.
Three basic cases,
according to the orientations of the NSs' magnetic moments with respect to the orbital angular moment of the BNS,
are discussed in producing Poynting-flux.
By assuming the fast magnetosonic wave is responsible for the radio emission,
the minimum magnetic field of NSs in pre-merger BNSs
for explaining the observed periods and luminosity of FRB~20191221A and FRB~20210213A
are estimated, i.e., $\sim8\times10^{12}\,\rm{Gs}$ and $\sim2.4\times10^{12}\,\rm{Gs}$, respectively.
Thus, we conclude that the minimum magnetic fields in BNSs mergers required for producing the one-off period FRBs like these bursts should be as high as $\gtrsim10^{12}\,\rm{Gs}$.

The neutron stars with high magnetic field ($\gtrsim 10^{12}\,\rm{Gs}$) are relatively rare
in the observed BNSs of our Galaxy,
most of which are consist of two low-magnetized NSs\footnote{\href{https://www.atnf.csiro.au/research/pulsar/psrcat/}{https://www.atnf.csiro.au/research/pulsar/psrcat/}} ($\sim10^{9}\,\rm{Gs}$; \citealp{ATNF-2005AJ}).
It means that
the BNSs being responsible for the one-off FRBs should be the other systems rather than the observed BNSs of our Galaxy,
which would merge with low magnetic fields finally.
In order to obtain a credible event rate of BNS-mergers related to their final magnetic fields,
we consider the evolution of both the BNS and their magnetic fields.
Based on the population synthesis and adopting a decaying magnetic field of NSs,
we estimate the event rate of BNS-mergers relative to their final magnetic fields.
It is found that the rapid merged BNSs,
of which the merger time is generally less than the decay timescale of the magnetic field in a pulsar (or magnetar),
tend to merge with high magnetization.
In the rapid merged BNSs, the companion will merge with the nearly initial magnetic field,
while the magnetization of the primary NS is often lower,
which makes them become an ideal energy reservoir for the pre-merger electromagnetic counterparts like FRBs.
In Milky Way, a rapidly merging BNS population is required for describing the observed heavy element abundances (\citealp{2019MNRAS.487.1426B, Hotokezaka-2018}), i,e., at least 40\,\% of the entire BNS population should merge within $1\,{\rm Gyr}$.
%\citealp{Hotokezaka-2018}
%A low-metallicity environment could produce more BH (black hole)-BH (or BH-NS) mergers and has a few increases in NS-NS mergers if the formation star is fixed.
%(\citealp{2010ApJ...715L.138B, 2010ApJ...714.1217B}).
%The BNS systems born in a low-metallicity environment have a shorter orbit (thus a shorter merger time).
%For a constant formation star rate, A low-metallicity environment could produce more BH (black hole)-BH (or BH-NS) mergers while having a few increases in NS-NS mergers.

The possible effect of accretion-induced magnetic field decay during the stages of mass transfer is not considered in this paper (\citealp{1974SvA....18..217B, 1994MNRAS.269..455J, 1997MNRAS.284..311K, 1999MNRAS.308..795K, 2001ApJ...557..958C, 2002MNRAS.332..933C, 2004MNRAS.348..661K, 2005ApJ...625..957L, 2006MNRAS.366..137Z}).
The mass transfer from a stripped post-helium-burning donor star (case BB) onto a NS is likely occurring in a dominant channel to ensure the NS' recycling happens (\citealp{2018MNRAS.481.4009V, 2020MNRAS.494.1587C}).
\cite{2015MNRAS.451.2123T} estimated the amount of accreted mass $\Delta{M}_{\rm NS}$ by the NS during the case BB in different orbital periods and helium star masses, and $\Delta{M}_{\rm NS}= 5\times10^{-5}-3\times10^{-3}\,M_{\odot}$ was obtained.
%by adopting the Eddington limit ($\overset{\cdot}{M}_{{\rm Edd}} \simeq 3\times10^{-8}\,M_{\odot}\, {\rm yr}^{-1}$)
%The recently observed magnetized accretion in compact binary systems supported their estimation ($\overset{\cdot}{M} \simeq 3\times10^{-9}\,M_{\odot}\, {\rm yr}^{-1}$; \citealp{2023arXiv230212318M}).
The accreted mass in case BB is significantly low compared with the magnetic-field-decay mass-scale ($\Delta{M}_{\rm decay} \simeq 0.01-0.02\,M_{\odot}$;\,\citealp{2006MNRAS.366..137Z, 2011MNRAS.413..461O}).
That is to say, the accretion of the primary NS in case BB has negligible effect on its magnetic field.
In the same reason, the effect of the accreted mass in the high-mass X-ray binaries ($\Delta{M}_{\rm NS}= {\rm a\;few} \times10^{-3}\,M_{\odot}$; \citealp{2017ApJ...846..170T, 2015A&A...577A.130F}) or NS-helium star stages ($\Delta{M}_{\rm NS}< 4 \times10^{-4}\,M_{\odot}$; \citealp{2017ApJ...846..170T}) on the suppression of the magnetic field in the primary NS can also be neglected.
The detailed evolution of the common envelope phase is still unclear (see \citealp{2023arXiv230308997B, 2022arXiv221207308R, 2013A&ARv..21...59I}, for reviews).
\cite{2017ApJ...846..170T} studied the recycling pulsar in Galaxy and
found that the mass transfer onto the primary NS during the common envelope is at most $0.02\,M_{\odot}$.
If all of the primary NS have accreted $\sim 0.02\,M_{\odot}$ mass and takes the magnetic field decay mass-scale $\Delta{M}_{\rm decay}=0.02\,M_{\odot}$ (see the Section 4.2 of \cite{2017ApJ...846..170T}),
the BNS-mergers rate with both NSs' magnetic field being higher than $10^{12}\,\rm{Gs}$
is reduced to $3.5\times10^{5}\,{\rm yr}^{-1}$ ($\sim 9\,\%$ of the total BNS-mergers) in redshift $z<1$.
However, the BNS-mergers rate with one NS's final magnetic field being higher than $10^{12}\,\rm{Gs}$
is not affected.

Some studies (\citealp{2020MNRAS.494.1587C, 2021MNRAS.504.3682C, szary2014radio, faucher2006birth}) found that
the decay timescale of the magnetic field can be up to $500-1000\,{\rm Myr}$, which means that $77-83\,\%$ of the total BNS-mergers will merge with their initial magnetic fields within $z=1$ in our calculation.
Such a high rate of BNS-mergers with strong magnetic fields may be challenged
by the lack of direct detection of the per-merger electromagnetic counterparts.
The pre-merger electromagnetic counterparts (e.g., X-ray/radio emission) of the BNS-mergers could be detected as a precursor of short gamma-ray bursts (\citealp{Hansen-2001MNRAS, 2013PASJ...65L..12T, Palenzuela-2013PhRvL, Most-2020ApJ}).
\cite{2010ApJ...723.1711T} revealed that around $8-10\,\%$ of the short gamma-ray bursts are accompanied with a precursor.
\cite{2020ApJ...902L..42W} performed a stringent search ($\gtrsim 4.5\sigma$ of the signal significance)
on for the precursor of short gamma-ray bursts and found the rate of $\sim3\,\%$ for bursts with a precursor.

The periodic radio emission as a precursor of BNS mergers is still undetected yet.
The event rate of BNS-merger-originated FRBs, i.e., BNS-mergers with both NSs' magnetic field being higher than $10^{12}\,\rm Gs$,
is $\sim8\times10^{4}\,\rm{yr}^{-1}$ (19\,\% of the total BNS-mergers) in redshift $z<1$.
Our estimation about the BNS-mergers shows that
nearly $19\,\%$ of the BNSs could merge with high magnetic fields ($\ge 10^{12}\,\rm{Gs}$) for both the primary NS and the companion NS,
which implies that
nearly one of five detected gravitational-wave events from BNS-mergers
could produce the one-off radio signal (like FRB~20191221A and FRB~20210213A)
if the beaming angle of the radio emission is not considered.

\acknowledgments
This work is supported by the National
Natural Science Foundation of China (grant Nos. 12273005,
11673006, U1938116, U1938201, U1731239, and U1938106),
the Guangxi Science Foundation (grant Nos. 2018GXNSFFA281010,
2017AD22006, 2018GXNSFGA281007, and 2018GXNSFDA281033), and China Manned Spaced Project (CMS-CSST-2021-B11).

\end{document}